\documentclass[manuscript]{emulateapj}
\usepackage[dvips]{color} 

\newcommand{\mjybm}{\mbox{mJy~beam${}^{-1}$}}
\newcommand{\sci}{Science}
\newcommand{\atel}{Astron.~Telegr.}

\slugcomment{Not to appear in Nonlearned J., 45.}

\shorttitle{Position wander of the radio jet base}
\shortauthors{Niinuma et al.}

\begin{document}


\title{Discovery of a wandering radio jet base after a large X-ray flare in the blazar Markarian 421}


\author{K. Niinuma\altaffilmark{1}, M. Kino\altaffilmark{2}, A. Doi\altaffilmark{3}, K. Hada\altaffilmark{4}, H. Nagai\altaffilmark{5} and S. Koyama\altaffilmark{6}}

\altaffiltext{1}{Graduate School of Science and Engineering, Yamaguchi University, Yoshida 1677-1, Yamaguchi, Yamaguchi 753-8512, Japan}
\altaffiltext{2}{Korean VLBI Network, Korea Astronomy and Space Science Institute, Daedeokdae-ro 776, Yuseong-gu, Daejeon 305-348, Republic of Korea}
\altaffiltext{3}{The Institute of Space and Astronautical Science, Japan Aerospace Exploration Agency, 3-1-1 Yoshinodai, Chuou-ku, Sagamihara, Kanagawa 229-8510, Japan}
\altaffiltext{4}{Mizusawa VLBI Observatory, National Astronomical Observatory of Japan, 2-21-1 Osawa, Mitaka, Tokyo 181-8588, Japan}
\altaffiltext{5}{Chile Observatory, National Astronomical Observatory of Japan, 2-21-1 Osawa, Mitaka, Tokyo 181-8588, Japan}
\altaffiltext{6}{Max-Planck-Institut f\"{u}r Radioastronomie, Auf dem H\"{u}gel 69, D-53121 Bonn, Germany}

\begin{abstract}
We investigate the location of the radio jet bases (``radio cores'') of blazars in radio images, 
and their stationarity by means of dense very long baseline interferometry (VLBI) observations. 
In order to measure the position of a radio core, we conducted 12 epoch astrometric observation of the blazar 
Markarian 421 with the VLBI Exploration of Radio Astrometry at 22 GHz immediately 
after a large X-ray flare, which occurred in the middle of 2011 September. 
For the first time, we find that the radio core is not stationary but rather changes its location toward 0.5 mas 
downstream. This angular scale corresponds to the de-projected length of a scale of $10^5$ Schwarzschild 
radii ($R_\mathrm{s}$) at the distance of Markarian~421. 
This radio-core wandering may be a new type of manifestation associated with the phenomena of large X-ray flares.
\end{abstract}


\section{Introduction}
Blazars are extremely bright active galactic nuclei (AGNs) with variable jets. Blazar jets are 
thought to be powered by the accretion of  surrounding material onto super-massive black holes (``central engines"). 
The radio jet bases (``radio cores'') located in the upstream ends of AGN jets 
in radio images have been basically considered to be opaque photospheres 
(surfaces of an optical depth of $\sim1$) against synchrotron self-absorption 
(SSA) or free-free absorption. 
Frequency-dependent offsets of the radio core position detected by 
very long baseline interferometry (VLBI) observations support this 
characteristic structure \citep[e.g.,][]{marcaide83,hada11}.
On the other hand, 
the stationary standing shock model, which is also another idea explaining blazar radio 
cores has been proposed based on recent VLBI observations \citep[e.g.,][]{marscher08}. 
Although detailed radio observations can provide the morphological structure of an inner jet, 
their stationarity is less well understood in addition to the location of the radio core still being
 a matter of debate \citep[e.g., ][]{marscher08, hada11}. 
Long-period VLBI astrometry at low frequency has not only derived the stationarity of radio cores 
with an astrometric accuracy \citep[e.g.,][]{bartel86,guirado95,fey97} but also caught the sign of 
their wandering behavior \citep[][]{ros99,marti11}.
In particular, the stationarity of blazars related to high-energy activity has not yet been 
verified by observations within a limited period after the flaring phenomena.

The high-synchrotron-peaked BL Lac object Markarian~421, at redshift $z=0.031$, is known as 
the first bright $\gamma$-ray blazar in the TeV energy range to be reported by the Whipple ground-based 
Cherenkov telescope \citep{punch92}. This blazar harbors a $\sim 3.6 \times 10^8$~solar mass black hole \citep{wagner08}. 
\citet{niinuma12} have reported on superluminal apparent inward motion of the jet component relative to the radio core, 
which has steep spectra at more than $\sim$10 GHz \citep{lico12, blasi13}, 
using the Japanese VLBI network (JVN) soon after the largest X-ray flare occurred in 2010.
It is clear that inward motions along outflowing jets are not realistic, 
but these apparent motions can be explained if the position of the radio core changes \citep{kellermann04}.

To clarify the origin of such inward motions, 
we performed 12 epochs of astrometric observations of the large X-ray flare of Markarian~421 in 2011 September \citep{hiroi11} 
using the VLBI Exploration of Radio Astrometry (VERA) at 22 GHz together with a nearby 
reference source J1101+3904 (see Fig. \ref{fig:f1}). 
Throughout this paper, we adopt cosmological parameters of $H_0=70.2~\mathrm{km s^{-1}~Mpc^{-1}}$, $\Omega_\mathrm{\Lambda}=0.725$ \citep{komatsu11}. 
Based on these parameters, 1 mas at $z=0.031$ corresponds to a linear scale of 0.62 pc.

\section{Astrometric observation and data analysis}

\subsection{VERA observations}

We conducted 12 epochs of phase-referencing observations at 22GHz between 2011 September 16 
and 2012 April 28 with typical interval of 20days after the aforementioned large X-ray flare. 
Observations were conducted using VERA (baselines ranging between 1000 and 2300 km). 
VERA was designed for dedicated astrometric observations by installing a dual beam system (beams A, B) 
and is operated by the National Astronomical Observatory of Japan.
	
To measure the absolute position of Markarian~421, we selected J1101+3904, which is listed 
in the VLBA calibrator survey list\footnote{http://www.vlba.nrao.edu/astro/calib/vlbaCalib.txt}, as a position reference. 
%
%
\begin{figure}
\centering
\includegraphics[width=0.95\linewidth]{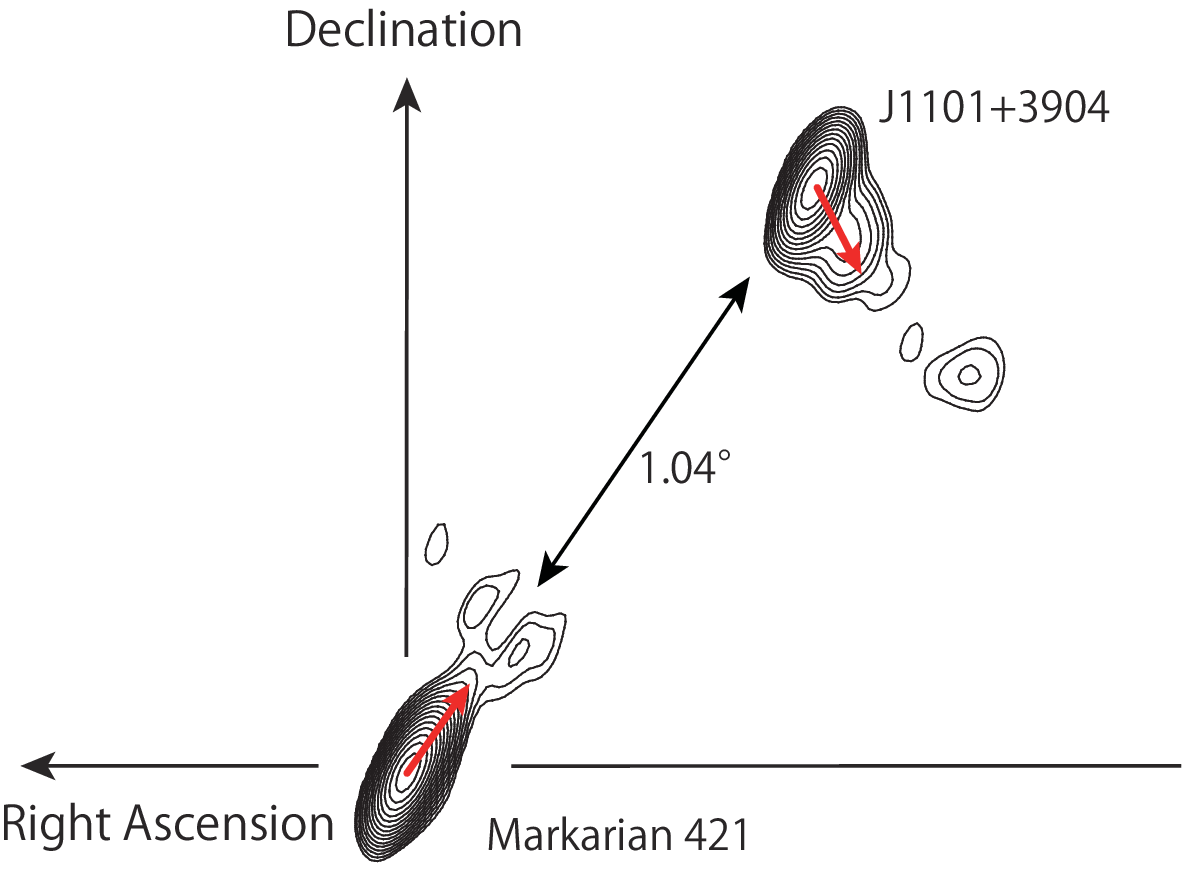}
\caption{Configuration of Markarian~421 and J1101+3904 jets on the celestial sphere. The tracking centers for both Markarian~421 and J1104+3904 are R.A. $=11^\mathrm{h}04^\mathrm{m}27^\mathrm{s}.313943$, decl. $=+38\arcdeg12\arcmin31\arcsec.79906$, and R.A. $=11^\mathrm{h}01^\mathrm{m}30^\mathrm{s}.069578$, decl. $=+39\arcdeg04\arcmin32\arcsec.63332$, respectively. The separation angle of both sources is 1.04$\arcdeg$. Red arrows represent jet orientations of both sources.} 
\label{fig:f1}
\end{figure}
In Fig. \ref{fig:f1}, we represent the configuration of both sources on the celestial sphere. 
To clearly display the source morphologies and jet orientations of Markarian 421 and J1101+3904, 
we show 8~GHz VLBI images of both sources observed by the Very Long Baseline Array (VLBA)\footnote{http://astrogeo.org/images/}.
The position angle (P.A.) measured from north to west of the jet of J1101+3904 is 
153\arcdeg and that of Markarian~421 is 35\arcdeg on average, respectively.
These jet orientations were derived from VERA images and are indicated with red arrows 
in Fig. \ref{fig:f1}. Therefore, the two jet axes are nearly perpendicular (118\arcdeg). 
In addition, the separation angle between two sources on the celestial sphere is only 1\arcdeg.04. 
The orthogonality and the closeness of two sources are crucial to avoid any degeneracy 
in the structural effects in the jet direction and to reduce the astrometric errors caused by the tropospheric delay residuals. 
Therefore, the selection of this pair of jets allows us to measure the absolute position of Markarian~421 with respect to 
J1101+3904 with high astrometric accuracy.

Our VERA observations were performed with a recording rate of 1024Mbps (2-bit quantization $\times$ 16~MHz $\times$ 
16 intermediate frequency (IF) channels ). Eight IF channels were assigned to Markarian~421 (beam A), 
and the others were assigned to J1101+3904 (beam B). 
The instrumental phase difference between the two beams was monitored in real time during the observations 
by injecting artificial noise sources into both beams at each station \citep{honma08a}.

\subsection{Data analysis}
The visibility calibration was performed using the NRAO Astronomical Image Processing System (AIPS) software package.
To increase astrometric precision, we recalculated the delays of all sources using a precise geodetic model and calibrated 
the instrumental phase difference between the dual beam systems. This precise recalculation of the delay also included 
the most recent Earth rotation parameters provided by the International Earth Rotation and Reference System Service (IERS)
 and the tropospheric delays measured using global positioning system receivers at each VERA station \citep{honma08b}. 
 Ionospheric delays were considered based on a global ionosphere map, which was produced every 2hr by the University of Bern.

\begin{figure}
\centering
\includegraphics[width=1\linewidth]{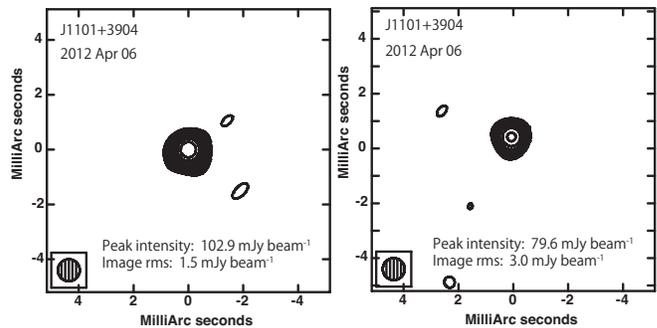}
\caption{\textsl{Left} and \textsl{right} panels show, respectively, the self-calibrated and phase-referenced VERA images 
of J1101+3904 on 2012 April 6.
These images are convolved with a circular Gaussian function that has a full-width at half-maximum of 0.8 mas, 
the typical minor-axis size of the synthesized beam in our observations, shown in the bottom-left corner.} 
\label{fig:f2}
\end{figure}

We performed fringe-fitting on both Markarian~421 and J1101+3904 independently using the AIPS task FRING and removed 
the residual delays, rates, and phases assuming a point-source model. Then VLBI images of Markarian~421 and J1101+3904 
(Fig. \ref{fig:f2} \textsl{left}) were produced via a standard CLEAN and self-calibration procedure using the Caltech Difmap 
package \citep{shepherd97}.

To obtain the core position of Markarian~421, we performed a phase-referencing analysis in the following manner. 
First, we chose Markarian~421 as the phase calibrator because this source is much brighter than J1101+3904. 
After the a priori corrections of the visibility amplitude, the solutions for residual delays, rates, and phases obtained 
on Markarian 421 (under the point-source assumption) were transferred to the data set of J1101+3904. 
Then, to account for the source structure phase of Markarian~421, we ran the AIPS task CALIB on Markarian~421 
with the source model derived from the self-calibrated image, and the solution was also transferred to J1101+3904 
for accurate determination of the absolute position of the radio core. Based on these procedures, we finally obtained 
phase-referenced images of J1101+3904 (Fig. \ref{fig:f2} \textsl{right}), which displayed the relative position of 
J1101+3904 with respect to Markarian~421. In the astrometry result, we present the relative position of Markarian~421 
with respect to J1101+3904.

To determine the jet orientation of both sources utilized through this paper, 
we fitted two-dimensional circular Gaussian models to these self-calibrated visibilities. 
The jet orientation of Markarian 421 derived from the VERA observations is consistent with that reported by previous 
VLBA observation \citep{lico12}. The P.A. of the innermost jet in J1101+3904 ($\sim$0.8 mas from the core) derived from 
VLBA 8 GHz data (see Fig. \ref{fig:f1}) is approximately 150\arcdeg, which is also quite consistent
 with VERA observations. 

The astrometric accuracy in each observing epoch represented in Fig. \ref{fig:f3}-\textbf{(b)} and \ref{fig:f4} is 0.08--0.12 mas. 
To estimate these values, we calculated 
(1) the statistical astrometric error derived from the signal-to-noise ratio of the phase-referenced image of J1101+3904, 
(2) the dispersive ionospheric residuals, (3) the non-dispersive tropospheric residuals, and (4) the identification of the core position 
by following the way described in \citet{hada11}.

\section{Detection of the position wandering of the radio jet base}\label{sect3}
%
\begin{figure}
\centerline{\includegraphics[width=\linewidth]{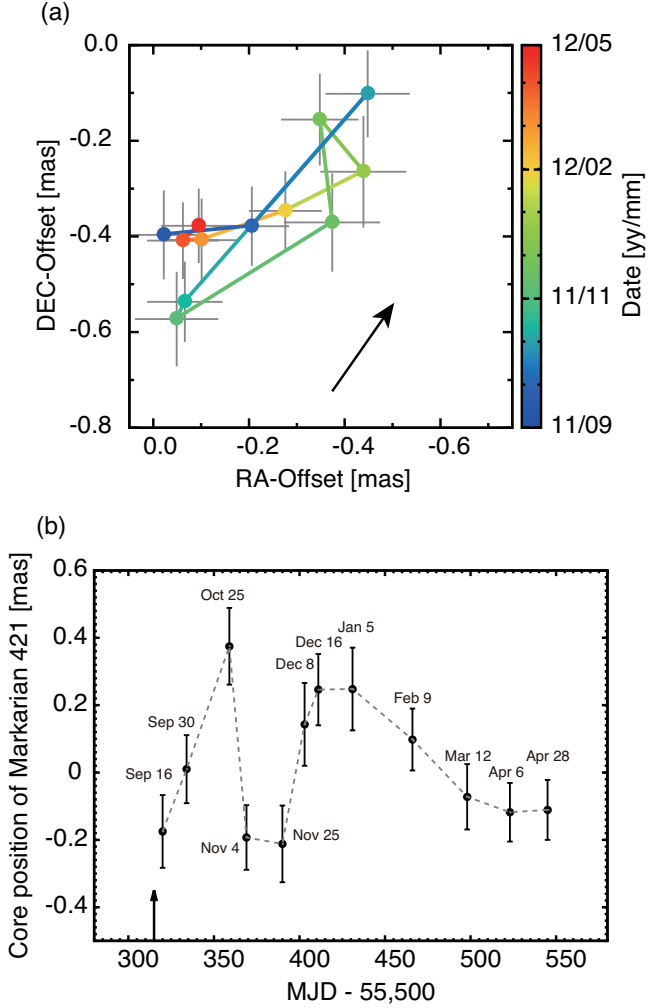}}
\caption{Measured radio-core position of Markarian~421 with respect to the position of the reference source J1101+3904. \textbf{(a):} The horizontal and vertical axes and the color gradient represent the positional offsets of right ascension and declination from the observation coordinates with one-sigma positional uncertainties and the observation date, respectively. The black arrow indicates the jet direction of Markarian~421. 
\textbf{(b):} The horizontal and vertical axes show the time and the positional offsets 
along the perpendicular direction to the jet of J1101+3904. 
The black arrow indicates the date on which the large X-ray flare occurred.}
\label{fig:f3}
\end{figure}
We discovered a core-position wandering in Markarian~421 based on bonafide astrometric observations (Fig. \ref{fig:f3}). 
This is the first direct detection of such a clear position change of the radio core 
within a year at higher frequency ($>~20$ GHz).
We show the position of the Markarian~421 radio core in the celestial coordinates 
in Fig. \ref{fig:f3}-\textbf{(a)}. 
The core clearly shows its position change in the northwest direction, 
which is a similar direction to the Markarian~421 jet axis. 
In Fig. \ref{fig:f3}-{\bf (b)}, we present the measured core positions of Markarian~421 projected 
to the P.A. of 63\arcdeg, which is a direction perpendicular to the J1101+3904 jet axis, 
because the structural effect of J1101+3904 is negligible in the motion along this direction. 
The position of the radio core began to change significantly $\sim20$ days after the peak of the X-ray flare. 
The radio core, which is located at $-0.16\pm0.11$ mas on 2011 September 16 
(modified Julian date (MJD) 55820) in Fig. \ref{fig:f3}-\textbf{(b)}, 
moved in the positive direction to $0.35\pm0.11$ mas by 2011 October 25 (MJD 55859). 
The core position changed by 0.51 mas within 39days, which corresponds to an apparent velocity of $\sim9~c$. 
After the core returned to near the initial position on MJD 55890, a core-position change of nearly the same magnitude, 
0.49 mas, was detected again within the 41 days between MJD 55890 and MJD 55931. The radio core then gradually 
returned once more to its initial position by approximately MJD 56000. 
We stress that the amount of core-position wandering is significantly larger than our astrometric accuracy of 
$\sim100~\mu\mathrm{as}$.

For quantitative evaluation of the significance of the radio core wandering during all the observing epochs, 
we examine the chi-square test for the position change represented in Fig. \ref{fig:f3}-\textbf{(b)}.
In this test, we define two cases of null hypothesis as follows: the position of the radio core shows the stationarity for 
1) the weighted mean position of the radio core during all the observing epochs and 2) the weighted mean position 
of the radio core during last three epochs in which it seems to be stationary.
In both cases, the stationarity of the radio core is rejected at a significance level of $< 0.1\%$ as shown in Table \ref{tbl:t1}. 
The radio core, hence, exhibits its position change during observing epochs at a confidence level of $> 99.9\%$.

%
\begin{figure*}[htbp]
\centerline{\includegraphics[width=\linewidth]{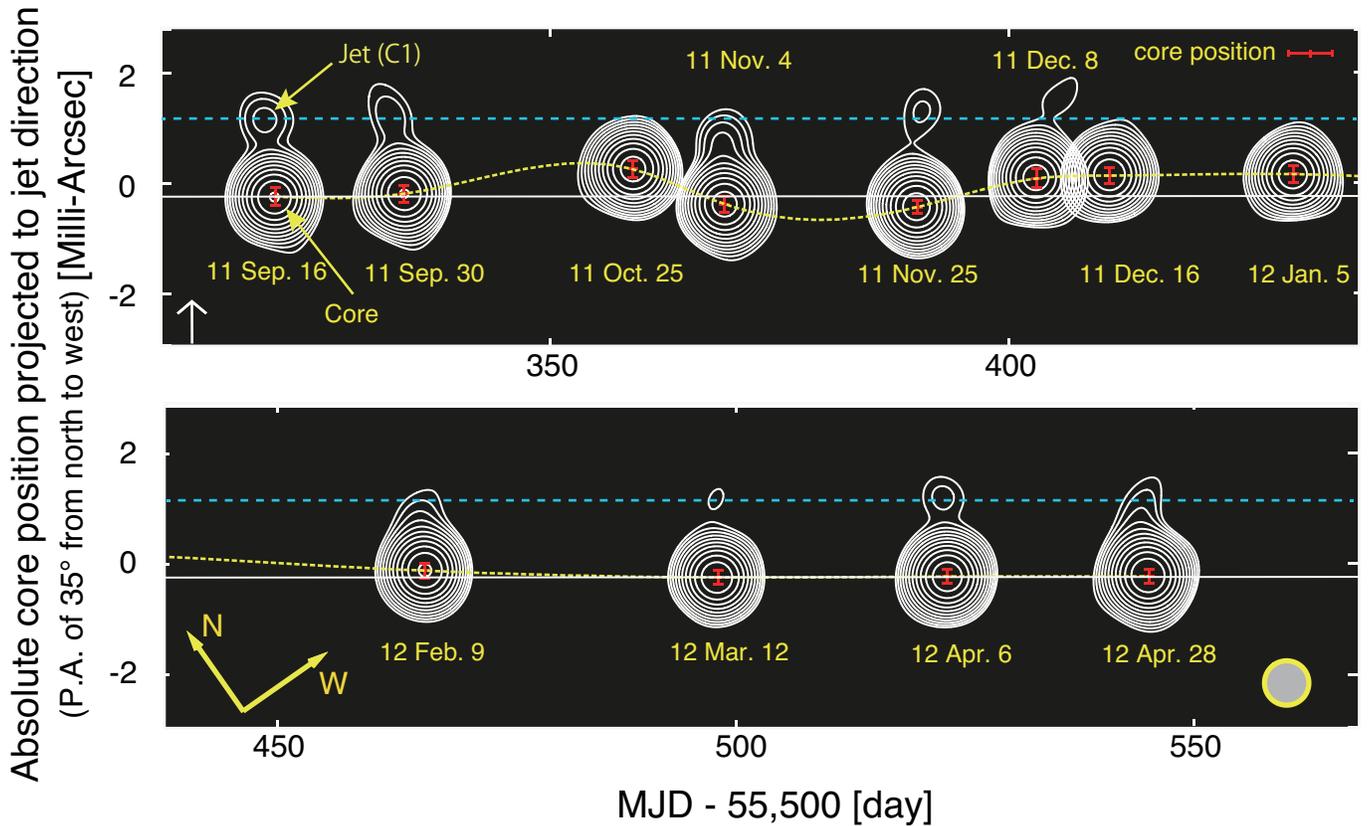}}
\caption{Radio-core positions projected on the jet axis and radio images of Markarian~421. These images are overlaid on their corresponding absolute core positions and were convolved with a circular Gaussian function that has a full width at half maximum of 0.8 mas (shown in the bottom-right corner). The contours begin at 6 \mjybm~and increase in steps of $2^{n/2}$. The vertical axis represents the jet axis. The white arrow in 2011 September indicates the date of the large X-ray flare. The white solid and the blue dashed lines indicate the positions of the radio core and the jet component (C1) as measured in 2011 September 16, respectively. The yellow dashed line represents the cubic spline interpolation of each core position.}
\label{fig:f4}
\end{figure*}

In Fig. \ref{fig:f4}, we present the absolute core positions along the jet axis of Markarian~421, and radio images are 
overlaid on each corresponding epoch. In this figure, the presence of a single jet component (hereafter, we call it `C1') 
located at approximately 1.3 mas northwest from the core can be confirmed and is also well known as its stationarity 
\citep{lico12}. C1 is much fainter than the radio core as represented in Fig. \ref{fig:f4}. 
In three out of twelve epochs (namely 2011 October 25, 2011 December 16, and 2012 January 05), the radio core significantly moves 
downstream (in the northwest direction) of the jet, and the separations between the bright core and C1 become too close to 
resolve these two components with VERA's synthesized beam. 
Hence, C1 was not seen in these three epochs. Thus, the 12 epoch sequence of the radio images suggests the core 
position wandering phenomenon discovered by our astrometric observations.

Moreover, 
if a displacement of the brightness centroid between the stationary core and a newborn jet 
causes the core wandering phenomenon, 
an elongated or double-peaked radio core image around the reference stationary position (the white line), 
which should be expected in the epochs where the core clearly moves downstream of the jet in Fig. \ref{fig:f4} 
\citep[e.g., radio images with a beam size of 0.1 mas in][]{marscher10}. 
However, such signatures were not observed in the radio-core images. Therefore, the wandering behavior of the 
radio core discovered in the present study clearly differs from the observational feature of the radio core in, e.g., 
BL Lacertae, in which the radio core is indicated to be stationary \citep{marscher08}.

%
\begin{table}[htbp]
\begin{center}
\caption{Chi--square test for stationarity of the radio core position.\label{tbl:t1}}
\begin{tabular}{lcc}
\tableline\tableline
   Parameters       &  Case (1)     &  Case (2)    \\
\tableline
     $\mu^{\dagger}$ &  $0.002$  &  $-0.106$ \\
     $\chi^2 / \nu^{\ddagger}$ & $40.99 / 11$ & $56.00 / 11$\\
     $P_{\mathrm{stationary}}(\chi^2; \nu)^{*}$ & $< 0.1$  &  $< 0.1$   \\
\tableline
\end{tabular}
\tablenotetext{}{\textbf{Note.}}
\tablenotetext{$\dagger$}{$\mu$ is the weighted average position of the radio core projected to the P.A. of 63\arcdeg in mas.}
\tablenotetext{$\ddagger$}{$\nu$ is the number of degrees of freedom.}
\tablenotetext{*}{Probability (\%) that the radio core is stationary for the position $\mu$}
\end{center}
\end{table}

\section{Discussion}
The de-projected distance that corresponds to the core-position changes of Markarian~421 is approximately 
$3.5\sim8.7~\mathrm{pc}$ (equivalent to $(1.0~-~2.6)\times10^5$~Schwarzschild radii ($R_\mathrm{s}$)) 
for a jet-viewing angle of $2-5$\arcdeg \citep{lico12}. 
One of the leading models for explaining the blazar phenomenon is the so-called internal-shock model, 
 in which discrete ejecta with higher speeds catch up with the preceding slower ejecta, and 
 the collision leads to a shock wave in the colliding ejecta \citep{spada01,guetta04}. 
Based on this model, core wandering can be naturally explained because the model predicts some 
scattering of the shock locations caused by the randomness of the ejecta speed. 
 Therefore, we assume that the distances of the shock locations from the black hole immediately 
 after the flare, $D\sim10^5~R_\mathrm{s}$, are proportional to the product of the square of the bulk Lorentz factor 
 ($\Gamma_\mathrm{jet}$) of the ejecta and the separation length ($\Delta$) of the colliding ejecta.
Assuming that a typical separation length is a scale of about 10 times the innermost stable circular orbit for 
non-rotating black holes ($r_\mathrm{isco}=3~R_\mathrm{s}$), in light of numerical simulations of relativistic outflows 
around black holes~\citep{mckinney13}, very fast ejecta with a Lorentz factor of approximately 
$\Gamma_\mathrm{jet}\sim60~(D/10^5~R_\mathrm{s})^{1/2} (\Delta/30~R_\mathrm{s})^{-1/2}$ 
are required to explain the observed magnitude of the core wandering. The $\Gamma_\mathrm{jet}$ is comparable to 
previously reported maximum values \citep[$\Gamma$ of $40\sim50$;][]{lister09,lister13} based on statistical 
studies by a long-term monitor using VLBI.
Moreover, the radio core returned to the reference stationary position after 2012 March 12 (the 10th epoch). 
This can be explained if the emission from the reference position is identical to the radio core that is in a quiescent state. 
The radio-core emission in a quiescent state is expected to be persistent.
As one of the possibilities for reproducing the observed core wandering, this persistent radio core was obscured by 
the flare-associated radio core, which therefore must be optically thick against SSA 
and short lived (a lifetime of approximately 2 weeks), for several epochs. Such a flare-associated radio core 
must be located at least $10^5~R_\mathrm{s}$ downstream from the persistent radio core.
In addition, 
\citet{koyama15} recently conducted multi-epoch astrometric observations of Markarian~501, 
which is also one of the most nearby TeV blazars by using VERA. 
However, they detected no significant positional changes of the radio core during its quiescent state. 
This fact, therefore, seems to support  the radio core wandering seen after the large X-ray flare.

Another possible scenario, which may lead to the radio-core wandering is the presence of changes in the electron density and 
magnetic field strength in the jet flow. Such changes might be triggered when magnetohydrodynamic waves 
propagate in the jet \citep{rees71,cohen14} with an apparent wave-pattern speed of approximately $9~c$. 
The propagation of electromagnetic waves in a charge-starved jet proposed by \citet{kirk11} might also be generated at 
the launching point of the jet, although an assumption of charge starvation may not be warranted because of 
the high number density of low-energy electrons \citep{kino08}. 
In addition, \citet{gomez97} have predicted that
the interaction between the steady standing shock and a propagating perturbation that contains shocked regions 
and is trailed by a rarefaction along the jet causes a positional shift of the brightness centroid 
with a superluminal speed comparable to the one reported in this Letter.

Regardless of the core-wandering mechanism, 
it is evident that our new observational finding offers crucial hints to resolving a key jet origin question 
(i.e., particle versus electromagnetic power) in the innermost region of the TeV blazar Markarian~421. 
In addition, 
recent radio and optical observations have suggested that monitoring the behavior of polarized emission on a short timescale 
is also a promising way to understand the origin of the jet activity in the innermost region of blazars \citep{lico14,itoh15}.

Thus far, as one of the few examples, multi-epoch astrometric observations have been performed for the FR-I-type 
radio galaxy M87 \citep{acciari09,hada14}. \citet{acciari09} noted that the position of the M87 
radio core at 43GHz moved by no more than $6~R_\mathrm{s}$ during the flare that occurred in 2008, while \citet{hada14} 
similarly showed a stable (less than $40~R_\mathrm{s}$) nature for the 43GHz core position during the flare in 2012. 
By contrast, we report here the detection of a large change in the position of a radio core.  
The de-projected distance corresponding to the core-position changes of Markarian 421, ($1.0~-~2.6)\times10^5~R_\mathrm{s}$, 
is approximately ten thousand times larger than the upper limit in the case of M87. 
According to the AGN unified model \citep{urry95}, both the blazar Markarian 421 and the radio galaxy M87 are classified as 
FR-I-type galaxies, and their jet power is approximately of $P_\mathrm{jet}\sim10^{37}$ W \citep{rieger12,potter13}. Therefore, 
the difference is most likely related to the jet-viewing angle and the optical depth structure rather than the jet power. 
The results reported by \citet{piner10} and \citet{hada13} imply that the jet has a transverse structure composed of 
an inner fast spine flow and a surrounding slow layer \citep{ghisellini05}. On the basis of this model, observers see 
the radio core in the blazar, which is dominated by the fast spine flow because the jet is closely aligned with our line of sight. 
In contrast, the radio galaxy has a jet off-axis whose emission comes from the slow layer and not from inner spine flow. 
A structured jet is a possible model for explaining the observational discrepancy in the behavior of the radio cores between 
Markarian 421 and M87.

\section{Summary}
We conclude that the radio core-wandering phenomenon implies that 
the blazar emission zone in Markarian~421 is occasionally located at least $(1.0~-~2.6)\times10^5~R_\mathrm{s}$ 
downstream from the black hole just after the large X-ray flare. 
To wrap up, the location of the radio core in Markarian~421 is far from the black hole, as also suggested in other blazars 
but without performing astrometric observations \citep[e.g.,][]{marscher08,marscher10,abdo10,orienti13}, while the radio core is 
not stationary after the large X-ray flare. 
In addition, our result on the location of the radio core is consistent with the one in the case of CTA~102, which was 
classified as a flat spectrum radio quasar measured by the image alignment method \citep[][and references therein]{fromm15}. 
This radio core wandering was discovered just after the large X-ray flare phenomenon 
using dense VLBI astrometric observations. Thus there is a possibility that the wandering 
behavior may be a new type of manifestation associated with the phenomena of large X-ray flares.

\acknowledgments
We would like to thank T. Oyama, K. Fujisawa, K. Motogi, and T. Kawaguchi for discussions.
We also thank E. Ros for comments on the paper. 
VERA is operated by the National Astronomical Observatory of Japan. 
This work was partially supported by KAKENHI (24340042, A. D., and 24540240, M. K., 26800109, K. H.).



{\it Facilities:} \facility{VERA}

\clearpage

\end{document}